\def\beq{\begin{equation}}
 \def\eeq{\end{equation}}
 \def\beqa{\begin{eqnarray}}
 \def\eeqa{\end{eqnarray}}
 \def\beqa*{\begin{eqnarray*}}
 \def\eeqa*{\end{eqnarray*}}
 \def\barr{\begin{array}}
 \def\earr{\end{array}}

 \def\btabular{\begin{tabular}}
 \def\etabular{\end{tabular}}
 \def\btable{\begin{table}}
 \def\etable{\end{table}}
 \def\<{\langle}
 \def\>{\rangle}

\documentclass[aps,showpacs,showkeys,nofootinbib,prstab,preprint]{revtex4}
\usepackage{epsfig}
\usepackage{amsmath}
\usepackage{amsfonts}
\usepackage{amssymb}
\usepackage{graphics}
\usepackage{colordvi}
\begin{document}

\title{Open Cell Conducting Foams for High Synchrotron Radiation Beam Liners}

\author{ S. Petracca,\footnote{petracca@sa.infn.it} A. Stabile\footnote{arturo.stabile@sa.infn.it}}

\affiliation{Dipartimento di Ingegneria,
Universita' del Sannio, Palazzo Dell'Aquila Bosco Lucarelli, Corso Garibaldi, 107 - 82100, Benevento, Italy}
%
%--------------------------------------------------------------------------------------------------------------------------------------------------
\begin{abstract}
The possible use of open-cell conductive foams in high synchrotron radiation particle accelerator beam liners is considered. 
Available materials and modeling tools are reviewed,  potential pros and cons are discussed, and preliminary conclusions are drawn.
\end{abstract}
%--------------------------------------------------------------------------------------------------------------------------------------------------
%
% 29.20.db Storage rings and colliders 
% 29.27.-a Beams in PA 
% 41.60.Ap Synchrotron radsiation by moving charges
% 79.20.Hx secondary emission, 
% 88.40.fh Adv Materials development
%
\pacs{29.20.db, 41.60.Ap, 88.40.fh}
\keywords{Storage rings and colliders, Synchrotron radiation, Advanced Materials}
\maketitle
%
%---------------------------------------------------------------------------------------------------------------------------------------------------
\section{INTRODUCTION}
\label{sec:intro}
\begin{figure} [!h]
\centerline{ \includegraphics[width=5cm]{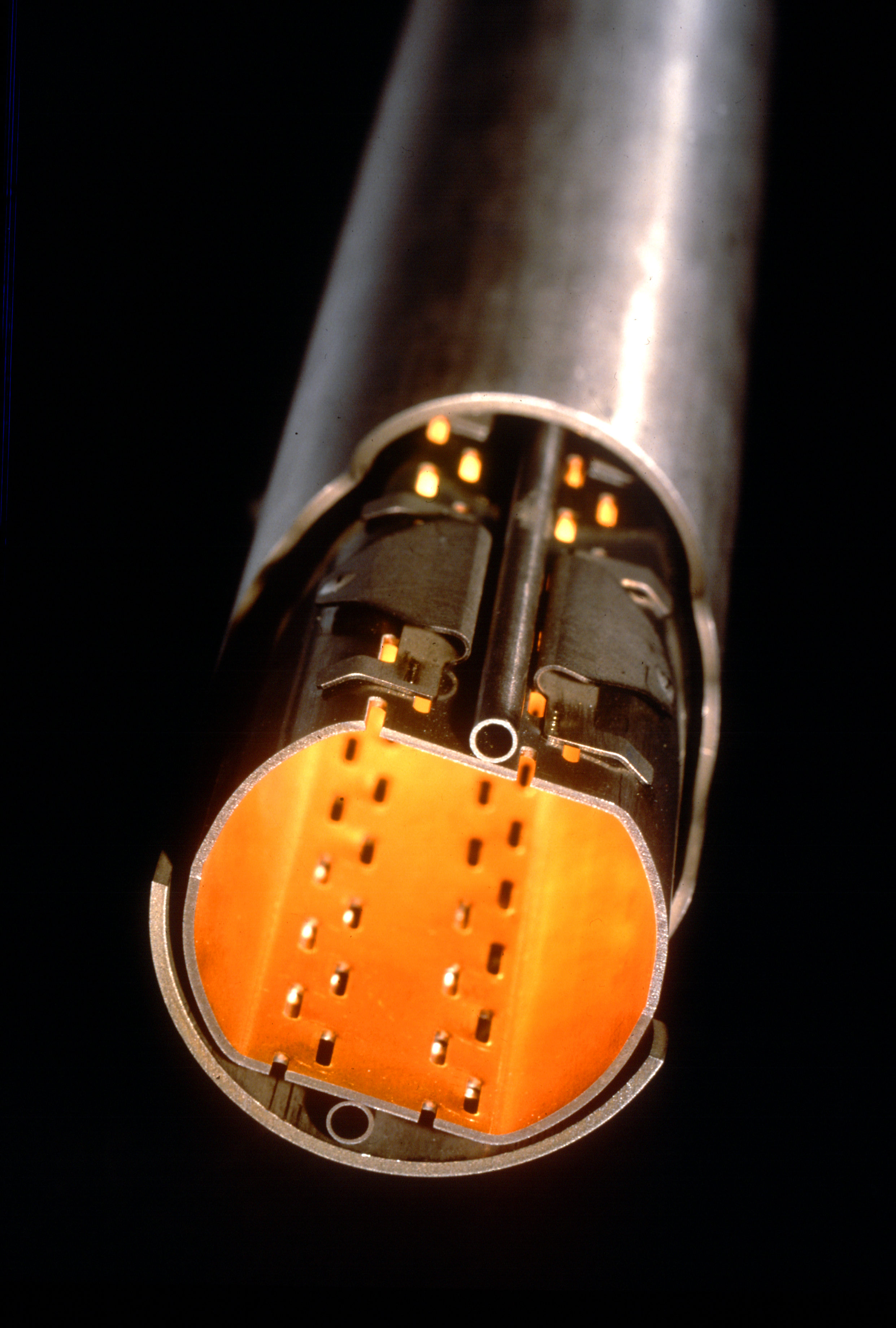}}
\caption{\it
The LHC slotted copper-plated beam pipe and stainless steel cold bore.}
\label{fig_1}
\end{figure}

%---------------------------------------------------------------------------------------------------------------------------------------------------
Molecular gas desorption from the beam-pipe wall 
due to synchrotron radiation
should be properly taken into account 
in the design of high energy particle accelerators and storage rings.
This has been a major (solved) challenge for the Large Hadron Collider;  
will be even more critical  for the HE-LHC, 
in view of its higher level of synchrotron radiation \cite{HELHC}, 
and will be a {\it crucial} issue for the successful operation 
of the proposed electron-positron Higgs factories \cite{epH}.\\ 
In the CERN Large Hadron Collider \cite{LHC} a copper-coated stainless-steel beam pipe (the {\it liner})
is kept at $\approx 20K$ by active Helium cooling, and effectively handles 
the heat load represented by synchrotron radiation, 
photoelectrons, and image-charge losses. 
A large number ($\sim 10^2$  $m^{-1}$) of tiny slots are drilled in the liner wall (see Figure \ref{fig_1})
in order to maintain the desorbed gas densities below a critical level
(e.g., $\sim 10^{15}$  $molecules/m^3$ for $H_2$) 
by allowing desorbed gas to be continuously cryopumped toward the 
stainless steel cold bore (co-axial to the liner) of the superconducting magnets,
which is kept at $1.9K$ by superfluid Helium.
Above such critical density levels, nuclear scattering in the residual gas would 
limit the beam luminositity lifetime,  eventually originate
high energy protons which may cause thermal runaway,
and ultimately cause quenching of the superconducting magnets.\\
%
%++++++++++++++++++++++++++++++++++++++++++++++++++++++++
%++++++++++++++++++++++++++++++++++++++++++++++++++++++++
%
The size, geometry, placement and density of the pumping slots 
affect the beam dynamics and stability  in a way which is
synthetically described by the longitudinal and transverse
beam coupling impedances \cite{Zotter}.  
The slot geometry and placement should be further chosen 
so as to minimize the effect of trapped (cut-off) modes, 
and  to prevent  the possible coherent buildup of  radiation 
in the TEM waveguide  limited by (the outer surface of) the pipe and the cold bore \cite{Heifets}.\\
Open-cell conducting foams  could be interesting candidate materials 
to help  fulfilling the above general requirements in 
beam liner design. 
In addition, the surface roughness of conducting foams may help reducing 
the effective secondary-emission yield (SEY) \cite{roughSEY}, 
thus alleviating the  electron-cloud build-up phenomenon \cite{ecloud},
and related instabilities \cite{ecloud2}.
In this  paper  we present  a brief review of  open cell conducting foams properties 
and modeling tools, 
and discuss at a very preliminary level the pros and cons of their possible use
in high synchrotron radiation accelerator beam liners. 
%
%+++++++++++++++++++++++++++++++++++++++++++++++++++++++++
\section{OPEN CELL METAL FOAMS}
\begin{figure} [!h]
\centerline{ \includegraphics[width=7cm]{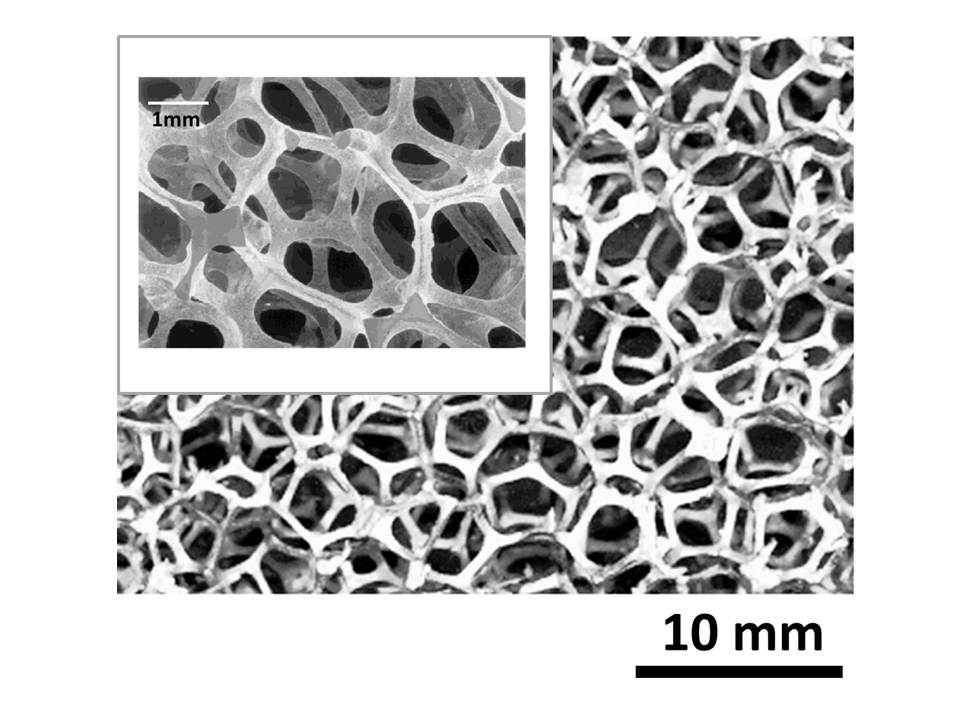}}
\caption{\it
A typical open cell metal foam, at two different viewing scales.}
\label{fig_2}
\end{figure}

\begin{figure} [!h]
\centerline{ \includegraphics[width=5cm]{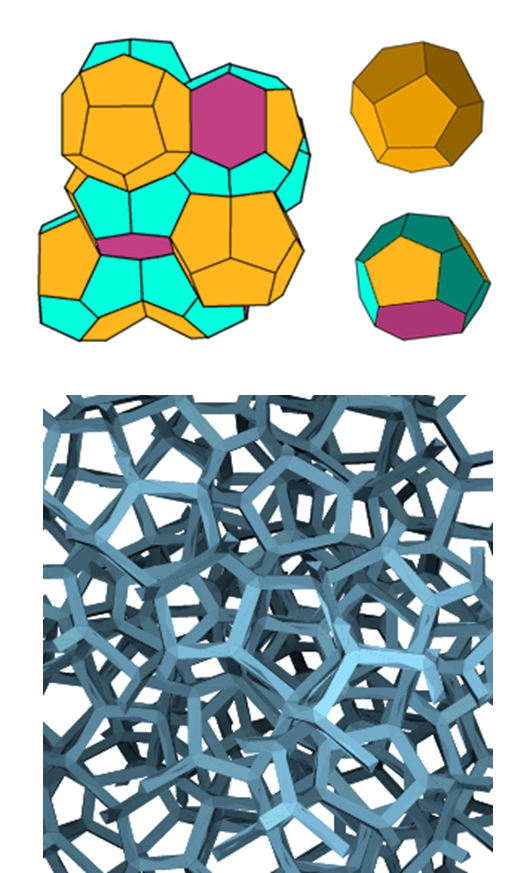}}
\caption{\it
The Wearie-Phelan honeycomb cell (top left), 
its constituent polyhedra (top right),
and  a numerically simulated reticulated foam thereof (bottom).}
\label{fig_3}
\end{figure}

%+++++++++++++++++++++++++++++++++++++++++++++++++++++++++
%
Open cell metal foams (OCMF) can be produced by vapor (or electro) deposition of metal 
on an open cell polymer template,  followed by polymer burn-off, and a final sintering step 
to densify the ligaments. 
Alternatively, they can be synthetised by infiltration/casting of molten metal into a solid mould, 
consisting of packed (non-permeable) templates of the foam pores, 
followed by burn-out and removal of the mould \cite{rev1}. 
Both processes result into highly gas-permeable {\it reticular} materials, 
where only a 3D web of thin conducting  ligaments survives.  
The typical structure of OCMFs is displayed in Figure \ref{fig_2}.
%
%++++++++++++++++++++++++++++++++++++++++++++++++++++++++
%++++++++++++++++++++++++++++++++++++++++++++++++++++++++
%
The key structural parameters of OCMF  are the "pore" size,
and the porosity (volume fraction of pores).
Pore sizes in the range from  $10^{-4}$ to $10^{-3}$ $m$ 
and porosities in the range  0.7 - 0.99  are typical.
These two parameters determine the  gas-permeability of the material, 
and, together with the electrical properties of the metal matrix, its electrical
characteristics.
OCMF have remarkable structural properties 
(low density and weight,  high (tensile and shear)-strength to weight ratio, 
nearly isotropic load response, low coefficient of thermal expansion), 
as summarized in Table I,
which  qualified  them among the most interesting new materials 
for aerospace applications.
%
%++++++++++++++++++++++++++++++++++++++++++++++++++++++++
%
\begin{table}[h]
\centering
\begin{tabular}{|l|c|c|c|}
\hline\hline
    &  Units & {\it Al} & {\it Cu} \\
\hline
Compressive Strength  & $[MPa]$  & $2.5$ & $0.9$  \\
\hline
Tensile Strength  & $[MPa]$  & $1.2$ & $6.9$  \\
\hline
Shear Strength  & $[MPa]$  &  $1.3$ & $1.3$  \\
\hline
Elastic Modulus (Compr.)  & $[MPa]$  & $1. \cdot 10^2$  &  $7.3 \cdot 10^2$  \\
\hline
Elastic Modulus (Tens.)  & $[MPa]$  & $1. \cdot 10^2$  &  $1. \cdot 10^2$  \\
\hline
Shear Modulus  & $[MPa]$  & $2. \cdot 10^2$  &  $2.8 \cdot 10^2$  \\
\hline
Specific Heat  & $[J/g^{0}C]$  & $0.89$  &  $0.38$  \\
\hline
Bulk Thermal Cond.  & $ [W/m^{0}C]$  & $5.8$  &  $10.1$  \\
\hline
Thermal Expansion Coeff.  & $[1/^{0}C]$  & $2.4 \cdot 10^{-5}$  &  $1.7 \cdot 10^{-5}$  \\
\hline
Bulk Resistivity  & $[ohm/m]$  & $7.2 \cdot 10^{-7}$  &  $6.5 \cdot 10^{-7}$  \\
\hline
Melting Point & $[^{0}C]$  & $660$  &  $1100$  \\
\hline\hline
\end{tabular}
\caption{ \it Structural properties of Al and Cu open cell metal foams from \cite{ERG}.}.
\label{table1}
\end{table}
%
%++++++++++++++++++++++++++++++++++++++++++++++++++++++++
%
Aluminum and Copper OCMF are presently available off-the-shelf 
from several Manufacturers worldwide, and are relatively cheap. 
They can be coated, with Silver, Titanium or Platinum, 
for special purpose  applications.
Steel and Brass foams, as well as Silver, Nickel, Cobalt, Rhodium, Titanium or Beryllium foams 
have been also produced.\\
The Weaire-Phelan (WP) space-filling honeycombs are credited 
to provide the  {\it natural}
(i.e., Plateau's  minimal surface principle compliant)
model  of  OCMF with equal-sized 
(but possibly unequal-shaped) pores \cite{WeaPh}.
The WP unit cell consists of a certain arrangement of (irregular) 
polyhedra, namely two pentagonal-face dodecahedra (with tetrahedral symmetry $T_h$) ,
and six tetrakaidecahedra (with antiprysmatic symmetry $D_{2d}$) featuring
two hexagonal and twelve pentagonal faces \cite{WeaPh}. 
A computer generated WP honeycomb is displayed in Figure \ref{fig_3}, and 
its visual similarity to Figure \ref{fig_2} is apparent.\\
A powerful open-source cross-platform software project for 3D morphological 
characterization  and modeling of  cellular materials, including OCMF, 
is under development \cite{iMorph}.\\
%
%++++++++++++++++++++++++++++++++++++++++++++++++++++++++
%++++++++++++++++++++++++++++++++++++++++++++++++++++++++
%
%++++++++++++++++++++++++++++++++++++++++++++++++++++++++
\subsection{Electrical Properties of Conducting Foams}
\begin{figure} [!h]
\centerline{ \includegraphics[width=10cm]{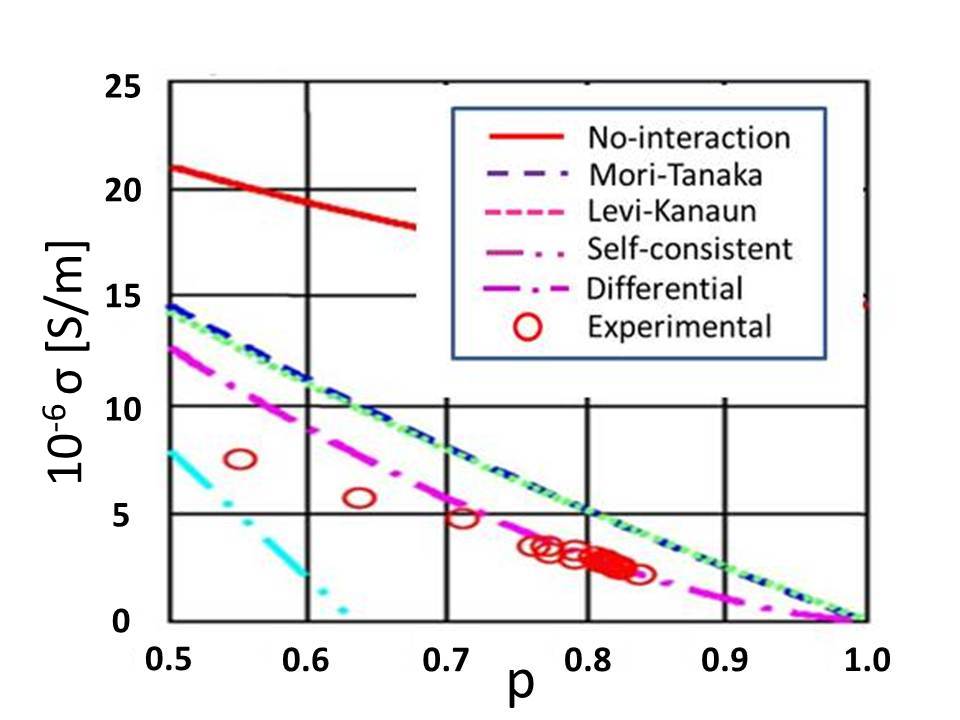}}
\caption{\it
Static conductivity vs porosity for Aluminum based foams from different
EMT models.  Measured values are also shown   
(adapted  from \cite{Sevostianov}).}
\label{fig_4}
\end{figure}

%++++++++++++++++++++++++++++++++++++++++++++++++++++++++
%
Electromagnetic modeling of OCMFs has been thoroughly investigated during the last decade.
A numerical approach based on Weiland finite integration technique
(FIT, \cite{Weiland}) has been proposed by Zhang et al. \cite{Zhang} to compute
the (frequency, thickness and angle of incidence dependent) reflection coefficient 
of  $SiC$  foam, and optimize its design.
The main limitation of Zhang's analysis is in the use of a simple body-centered-cubic
unit-cell foam model, for easiest numerical implementation. 
The FIT scheme, however may accommodate more complicated and realistic
foam-cell geometries, including in principle the WP one. \\ 
In the quasi-static limit $\lambda \longrightarrow 0$, 
the conductivity of  OCMFs can be computed using effective medium theory
(EMT), for which several formulations exist  (see, e.g., \cite{Rev1},\cite{RevN} for a review).
These include: 
i) the self-consistent approach \cite{effective_med}, credited to Bruggemann,
where inclusions are thought of as being embedded in the (actual) effective medium;
ii) the differential scheme, whereby inhomogeneities are {\it incrementally} added to the composite\footnote{
%--------------------------------------------------------------------------------
In this approach, the total concentration of inhomogeneities does {\it not} coincide with the volume fraction $p$, 
because at each step new inclusions may be placed where old inclusions have already been set.},
%-------------------------------------------------------------------------------- 
until the final concentration is reached, so that at each step the inclusions do not interact 
and do not modify the field computed at the previous step \cite{differential};
iii) the effective-field methods, whereby interaction among the inclusions is described in terms of an effective field 
acting on each inclusion, accounting for the presence of the others. 
Two main versions of this method exist, credited to Mori-Tanaka \cite{MorTan} and Levin-Kanaun \cite{LevKan}, 
differing in the way the effective field is computed 
(average over the matrix only, or average over the matrix {\it and} the inclusions, respectively).\\
The self-consistent approach yields 
\begin{equation}
\sigma_{\mbox{\it eff}}=\sigma_0(1-p\nu), 
\label{eq:ID}
\end{equation}
where $\sigma_0$ is the bulk metal conductivity, 
$p$ is the porosity (volume fraction of the vacuum bubbles),
and $\nu$ is a morphology-dependent factor.
The differential approach yields 
\begin{equation}
\sigma_{\mbox{\it eff}}=\sigma_0(1-p)^\nu,
\label{eq:diff}
\end{equation}
while the Mori-Tanaka/Levin-Kanaun approaches yield  
\begin{equation}
\sigma_{\mbox{\it eff}}=\sigma_0/(1+\frac{\nu p}{1-p}).
\label{eq:MT}
\end{equation}
All equations (\ref{eq:ID})-(\ref{eq:MT}) merge, as expected, in the $p \rightarrow 0$ limit.
The above models  are synthetically compared to
measured values of the static conductivity  in Figure \ref{fig_4},
for Aluminum foams.
%
%++++++++++++++++++++++++++++++++++++++++++++++++++++++++
%++++++++++++++++++++++++++++++++++++++++++++++++++++++++
%
For Aluminum foams, all these models  predict {\it larger} conductivity then observed in measurements\footnote{
%---------------------------------------------------------------------
It should be noted that open and closed cell metal foams behave similarly in terms of electrical conductivity,
while being markedly different as regards thermal conductivity, due to the different role of convective flow.}. 
%---------------------------------------------------------------------
This has been attributed to significant oxide formation on the solid web \cite{Clyne}.
Equation (\ref{eq:diff}) agrees in form with predictions 
based on percolation theory \cite{perco} - although strictly speaking
there is no threshold here beyond which the conducting component disconnects.\\
Measurements of the microwave electromagnetic shielding efficiency of OCMF panels 
\cite{Monti} indicate  that a simple Drude model
\beq
\sigma(\omega)=
\frac{\omega_p^2 \epsilon_0}{\imath \omega + \nu}
\eeq
provides a good description of the frequency-dependent conductance
of metallic foams.  Typically, the relevant plasma and collision frequencies, 
$\omega_p$ and $\nu$ are of the order of a few tens of GHz and
a few tens of KHz, respectively. These are notably much smaller\footnote{
%++++++++++++++++++++++++++++++++++++++++++++
In the light of the well known formula 
$$\omega_p^{2}=\frac{N_e q^2}{m_e\epsilon_0},$$ 
this can be explained  as due to a reduction 
of the  electron density $N_e$, roughly by a factor $(1-p)$, 
$p$ being the porosity and a parallel (large) increase
of the effective electron mass.  
Pendry et al. have shown that the effective mass of an electron 
in a thin-conducting-wire lattice or web 
is essentially set by its electromagnetic momentum, 
yielding \cite{Pendry}
$$
\omega_p^2 \sim \frac{c^2}{D^2 \log(D/d)},
$$
where $d$ and $D$ are the typical conducting-ligament  
and pore diameters, respectively, and
$$
\nu\sim\frac{\epsilon_0 \omega_p^2}{\pi \sigma_{0}}
\left(\frac{D}{d}\right)^2.
$$} .
%++++++++++++++++++++++++++++++++++++++++++++++++++++++++
%
than their solid metal counterparts (typically in the $PHz$ and $GHz$ range, respectively \cite{Ordal}).
%
%++++++++++++++++++++++++++++++++++++++++++++
\subsubsection{OCMF Impedance and Skin Depth}
\label{sec:Drude}
%++++++++++++++++++++++++++++++++++++++++++++
%
Throughout a typical beam current frequency spectrum,  
OCMF and bulk metals behave quite differently\footnote{
%
%+++++++++++++++++++++++++++++++++++++++++++++++
The typical power spectrum of a bunched beam in a ring collider consists of lines 
at integer multiples of $f_0=c/\delta_b$, $\delta_b$ being the bunch spacing,  
with an envelope approximately $\propto cos^2$.  In the LHC
the $-20$ dB  bandwidth is $\sim 1$ GHz, roughly   $10^5$  times 
the circulation frequency $\omega_R$, and $10^{-1}$ times
the cut-off frequency of the lowest waveguide mode of the beam pipe  \cite{Day}.}.
%+++++++++++++++++++++++++++++++++++++++++++++++
% 
In bulk metals , $\omega \ll \nu \ll \omega_p$  so that the  characteristic  impedance  $Z_m$ 
and (complex)  propagation constant $\tilde{k}_m  $ can be written 
\beq
Z_m  \sim 
\frac{1+\imath}{\sqrt{2}}
\left(
\frac{\omega\nu}{\omega_p^2}
\right)^{1/2}
Z_0,
\mbox{         }
\tilde{k}_m  \sim
\frac{1-\imath}{\sqrt{2}}
k_0(\omega_p)
\sqrt{\frac{\omega}{\nu}}
\label{eq:Zmetal}
\eeq
$Z_0$ and $k_0(\omega)=\omega/c$  being  the  vacuum characteristic impedance and propagation constant, respectively. It is seen that in metallic conductors, both  $Z_m$ and $\tilde{k}_m$  are  $\propto \omega^{1/2}$.\\
In OCMFs, on the other hand,   $\nu \ll \omega \ll \omega_p$  throughout the beam current spectrum,  
so that the material exhibits a {\it plasmonic}  behavior. The OCMF wall  (characteristic) impedance  $Z_{f}$  and (complex)  propagation constant  $\tilde{k}_{f}$ are thus given
(to lowest order in the small quantities $\nu/\omega$ and  $\omega/\omega_p$)   by
\beq
Z_{f} \sim  Z_0
\left(
\frac{\nu}{2\omega_p}+
\imath \frac{\omega}{\omega_p}
\right),
\mbox{         }
\tilde{k}_{f}  \sim
k_0(\omega_p)
\left(
\frac{\nu}{2\omega}-\imath
\right)
\label{eq:foamZk}
\eeq
Hence, the  OCMF  characteristic resistance  $R_{f}=\mbox{Re}[Z_{f}]$ 
and skin depth  $\delta_{f}$  are both {\it frequency independent},  
and,  e.g.,  for the case of high-grade 
($\rho\approx 5.5 10^{-10} \mbox{ ohm cm}^{-1}$ at $20K$) Copper foam with $p=0.9$,  
both fairly small:
\beq
R_{f} \sim \frac{Z_0}{2}\frac{\nu}{\omega_p} \approx 1.4 \cdot 10^{-5} \mbox{  ohm},
\mbox{       }
\delta_{f}\sim\frac{c}{\omega_p}\approx 6 \cdot 10^{-4} \mbox{ m}.
\label{eq:foam_delta}
\eeq
The OCMF characteristic reactance 
\beq
X_{f}=\mbox{Im}[Z_{f}] \sim \imath Z_0 \frac{\omega}{\omega_p}
\eeq 
on the other hand,  is large compared to that of bulk metal, 
and grows {\it linearly} with $\omega$.
For the case, e.g., of high-grade Copper foam with $p=0.9$, 
$X_{f}\approx 0.5 \mbox{ ohm}$ at $10^4 \mbox{Hz}$.\\
%
%+++++++++++++++++++++++++++++++++++++++++++++++
\subsection{Superconducting Foams}
%+++++++++++++++++++++++++++++++++++++++++++++++
%
At the operating  temperature of   the LHC liner ($\sim 20^oK$), 
both Aluminum and Copper exhibit a fairly large conductivity
($\sim 10^7 \mbox{ ohm$^{-1}$cm$^{-1}$} $),
but neither of them is superconducting.\\
Superconducting OCMFs have been discussed in \cite{SC_foams}. 
Foamed  {\it ceramic}  superconductors (in particular, YBCOs)
with critical temperature well above  $20K$  (HTS),
have also been already  manufactured  \cite{HTS_foams}. 
These materials may likely exhibit a very low SEY. \\ 
Besides being technologically appealing, HTS  foams
are conceptually interesting materials, 
where a percolating  electric current co-exists 
with a percolating magnetic flux \cite{Bartolome}. 
A substantial body of  experimental  results
on the electrical properties of thin-film HTS  is available \cite{HTS_EM_props}, 
and these are reasonably well accounted for \cite{2fluid_HTS} 
by a simple two-fluid Drude  model \cite{2fluid_Drude}.
However, the electrical properties of  {\it foamed} HTS
have been investigated sofar, 
to the best of our knowledge
mostly at very low frequencies \cite{HTS_foam_cond}.
%
%+++++++++++++++++++++++++++++++++++++++++++++++++++
\section{METAL FOAM vs  PERFORATED METAL  PATCHED PIPES}
\label{sec:vac}
%+++++++++++++++++++++++++++++++++++++++++++++++++++
%
In this section we shall attempt to draw a comparison 
between beam pipes using perforated-metal patches for outgassing
vs pipes using OCMF patches, 
in terms of the relevant vacuum 
and beam-coupling impedance features .
%
%+++++++++++++++++++++++++++++++++++++++++++++++++++
\subsection{Vacuum Issues}
\label{sec:vac1}
%+++++++++++++++++++++++++++++++++++++++++++++++++++
%
The vacuum dynamics for each molecular species which may be desorbed 
from the wall by synchrotron radiation can be described by
the following set of  (coupled) rate equations \cite{Grobner}:
\begin{equation}
\left\{
\begin{array}{l}
\displaystyle{V\frac{dn}{dt}=q-an+b\Theta}\\
\\
\displaystyle{F\frac{d\Theta}{dt}=cn-b\Theta}
\end{array}
\right.
\label{eq:rate}.
\end{equation}
Here $n \mbox{ }[m^{-3}]$ and $\Theta \mbox{ }[m^{-2}]$ are the
volume and surface densities of desorbed particles, respectively,
and $V$ and $F$ represent the volume and wall-area of the liner
per unit length, respectively.\\
The first term on the r.h.s. of the first rate equation represents the
number of molecules desorbed by synchrotron radiation per unit length
and time, and is given by
\begin{equation}
q=\eta\dot{\Gamma},
\end{equation}
where $\eta$ is the desorption yield (number of desorbed molecules per incident photon)
and $\dot{\Gamma}$ is the specific photon flux  
(number of photons hitting the wall per unit length and time). 
The second term 
%on the r.h.s. of the first rate equation 
represents the number of molecules which are removed 
per unit time and unit length 
by either sticking to the wall,
or escaping through the pumping holes/slots. 
The $a$ coefficient in (\ref{eq:rate}) can be accordingly written 
\begin{equation}
a=\frac{\langle v \rangle}{4}(s+f)F,
\label{eq:aa}
\end{equation}
where $\langle v \rangle\approx(8kT/\pi m)^{1/2}$ is the average molecular speed,
$m$ being the molecular mass, $k$ the Boltzmann constant and $T$ the absolute temperature, 
$\langle v \rangle/4$ is the average number of collisions of a single molecule
per unit time and unit wall surface, 
$s$ is the sticking probability,
and $f$ is the escape probability. 
%approximately equal to the fraction of wall surface occupied by the holes.
%
The third term  
%on the r.h.s. of the first rate equation
accounts for thermal or radiation induced re-cycling
of molecules sticking at the walls. 
The $b$ coefficient in (\ref{eq:rate}) can be accordingly written
\begin{equation}
b=\kappa\dot{\Gamma}+F \nu_o \exp(-W/kT).
\label{eq:bb}
\end{equation}
Here the first term accounts for radiation induced recycling, 
described by the coefficient $\kappa \mbox{ } [m^{2}]$,
while the second term describes thermally-activated recycling, 
$\nu_0$ being a typical molecular vibrational frequency, 
and $W$ a typical activation energy.\\
The $b\Theta$ term  appears
with reversed sign on the r.h.s. of the second rate equation,
where it represents the number of molecules {\it de-sticking} from the
wall surface per unit time and unit length.
The first term on the r.h.s. of this equation 
represents the number of molecules 
sticking to the wall, per unit time and unit length, 
whence  (compare with eq. (\ref{eq:aa})) 
\begin{equation}
c=\frac{\langle v \rangle}{4}s F.
\label{eq:cc}
\end{equation}
At equilibrium, $\dot{n}=\dot{\Theta}=0$, and the rate equations yield:
\begin{equation}
\left\{
\begin{array}{l}
\displaystyle{n_{eq}=\frac{4\eta\dot{\Gamma}}{\langle v \rangle f F}}\\
\\
\displaystyle{\Theta_{eq}=\frac{s}{f}
\frac{\eta\dot{\Gamma}}{\kappa \dot{\Gamma}+F\nu_o\exp(-W/kT)}
}
\end{array}
\right.
.
\label{eq:equil}
\end{equation} 
Typical values (for LHC) of the parameters in (\ref{eq:equil})
are collected in Table II below \cite{Grobner}.
%
%++++++++++++++++++++++++++++++++++++++++++++++++++++++++
%
\begin{table}[h]
\centering
\begin{tabular}{|c|c|c|}
\hline\hline
$V$    &  Liner volume (per unit length)  & $1.3 \cdot 10^{-3}$   $m^3/m$  \\
\hline
$F$    &  Liner surface (per unit length)  & $0.14$   $m^2/m$  \\
\hline
$\eta$ & Desorption yield & $5 \cdot 10^{-4}$ \\
\hline
$\dot{\Gamma}$ & Photon flux ($200mA$ beam) & $3.14 \cdot 10^{16}$  $s^{-1} m^{-1}$  \\
\hline
$s$   & Sticking probability & $0.6$  \\
\hline
$\kappa$  & Recycling coefficient  & $5. \cdot 10^{-21}$  $m^{2}$  \\
\hline
$\nu_0$   & Vibrational frequency  & $10^{13}$  $s^{-1}$  \\
\hline
$W$  & Activation energy  & $0.035$    $eV/molecule$  \\
\hline\hline
\end{tabular}
\caption{Typical values of the parameters in (\ref{eq:equil})  from \cite{Grobner}.}
\label{table2}
\end{table}
%
%++++++++++++++++++++++++++++++++++++++++++++++++++++++++
%
The equilibrium molecular densities in (\ref{eq:equil})  should not exceed some {\it critical} values
for safe operation \cite{Grobner}. 
%
%-------------------------------------------------------------------------------
\subsubsection{Liner with Perforated Metal Patches}
\begin{figure} [!h]
\centerline{ \includegraphics[width=16cm]{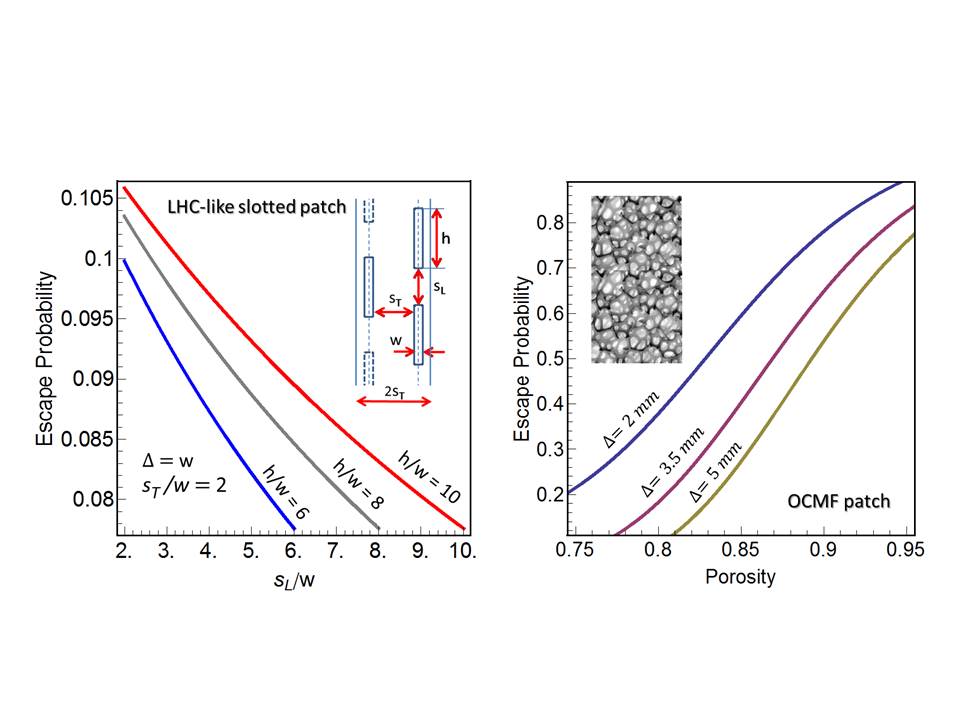}}
\caption{\it Left panel: escape probabilities for LHC-like slotted patches.  
Right panel: escape probability for OCMF patches.}
\label{fig_5}
\end{figure}

%-------------------------------------------------------------------------------
%
For a liner wall with vanishing thickness the molecular escape probability $f$ in
(\ref{eq:aa}) and (\ref{eq:equil}) will be given by the holey fraction $\xi_h$
of  the total wall surface. 
The desorption yield $\eta$ and the sticking probability $s$ 
will likewise differ from those of solid metal by a factor $(1-\xi_h)$.\\
For a thick perforated wall, the escape probability will be $f=\chi \xi_h$ 
where the factor $\chi < 1$ (named after Clausing), takes into account 
the possibility that molecules may stick at the hole {\it  internal} surface 
rather than escaping outside \cite{Steckel}.\\
If only a fraction $\xi_p$ of the liner surface is covered by perforated patches,
the liner escape probability will be $f=\xi_p \chi \xi_h$.\\
In the LHC  the pumping holes are thin slots parallel to the pipe axis \cite{LHCdes}. 
The slots are confined to four narrow ($\approx  9mm$ wide) strip-shaped patches, 
running parallel to the pipe axis, as seen in Figure 1.
The escape probability of an LHC-like slotted patch (strip) is  displayed 
in the left panel of  Figure \ref{fig_5}.
Stiffness requirements set a lower limit 
to the axial ($s_L$) and transverse ($s_T$) slot spacings at  
roughly twice the slot width $w$.
In figure \ref{fig_5} we assume $s_T=2w$ and   
plot the escape probability vs. the scaled longitudinal slot separation $s_L/2w$,
for three typical values of the slots height-to-width ratio $h/w$.
The wall thickness $\Delta$ is assumed as equal to $w$, yielding a
Clausing factor  $\chi \approx 0.68$ \cite{SteckMan}. \\
In the LHC the slotted strips represent roughly one half of the liner surface.
The slots are $\approx  8 mm$  long, on average, and $\approx 1.5 mm$ wide,
with $s_L \approx s_T \approx 3 mm$.  
The holey fraction of the total liner surface is roughly $5\%$, and the 
escape probability is $\approx 3.5\%$.
%
%
%++++++++++++++++++++++++++++++++++++++++++++++++++++++++
%++++++++++++++++++++++++++++++++++++++++++++++++++++++++
%
%------------------------------------------------------------------
\subsubsection{Liner with OCMF Patches}
%------------------------------------------------------------------
%
Gas permeability of OCMFs in the molecular flow regime 
is described by Knudsen diffusivity \cite{Steckel}. 
The dependence of this latter on foam porosity
was fitted numerically in \cite{Murch1} and \cite{Murch2},
using computer simulations of molecular diffusion
based on tomographic data of actual OCMF specimens.
An exponential blow up of Knudsen diffusivity  $D_K$  was observed,
above some porosity  threshold,  possibly 
due to the creation of long-range molecular pathways.\\
From Knudsen diffusivity the escape probability of OCMF patches 
can be readily obtained, using Fick's (diffusion) law, 
and is displayed in Figure \ref{fig_5}
(right panel) as a function of porosity, for different values 
of the OCMF wall thickness $\Delta$.\\
It is seen from Figure \ref{fig_5} that typical OCMF patches, 
e.g., with $p=0.9$,
feature fairly larger escape probabilities,
compared to slotted patches.\\
OCMF walls are also expected to provide better EM shielding 
compared to slotted ones.  Hence, gas molecules sticking inside the 
foam pores will be less exposed to recycling, resulting into possibly smaller values 
of the recycling factor $\kappa$ in (\ref{eq:bb}) and (\ref{eq:equil}).
%
%+++++++++++++++++++++++++++++++++++++++++++++++++++
\subsection{Beam Coupling Impedance and Parasitic Losses}
%+++++++++++++++++++++++++++++++++++++++++++++++++++
%
Beam coupling impedances provide a synthetic description
of the electromagnetic beam-pipe interaction.
Here, for the sake of brevity, 
we shall confine our discussion to the longitudinal impedance.  
The absolute value  and   the  imaginary part of this latter
are inversely proportional to the threshold currents 
for  (single-bunch) microwave instability
and Landau damping suppression, respectively,
and hence quite relevant to beam stability \cite{Zotter},
while the real part determines the parasitic loss  
(energy lost by the beam per unit pipe length),
via  \cite{Zotter}
\beq
\Delta{\cal E}=
   \frac{1} {2\pi}
   \int_{-\infty}^{+\infty }|I(\omega)|^2 
   \Re e~[\bar{Z}^{\parallel}(\omega) ]
   d\omega,
\label{eq:parloss}
\eeq
where $I(\omega)$ and $Z^{\parallel}(\omega)$
are the beam-current spectrum and longitudinal impedance per unit lenght,
respectively.\\
The following relationship exists
(to 1st order in the wall impedance) 
between the longitudinal impedance per unit length
$\bar{Z}_{\parallel}$ 
of a patched-wall  beam liner 
and the (known) longitudinal impedance  per unit length
$\bar{Z}^{(0)}_{\parallel}$ 
of the same pipe with a perfectly conducting wall \cite{Partacc}
\beq
\bar{Z}_{\parallel}\!=\!
\bar{Z}^{(0)}_{\parallel}+
\frac{\epsilon_0 Y_0}{c \Lambda^2}
\oint\limits_{\partial S} 
\!\! 
Z_{w}(s)
|E_{0n}(s)|^2
ds,
\label{eq:ZZ}
\eeq
where $Z_w$ is the (local)  Leont\'{o}vich impedance \cite{Leont}  of the patched wall,
$\partial S$ is the pipe cross-section contour,
$Y_0=(\epsilon_0/\mu_0)^{1/2}$ the vacuum admittance, 
$\Lambda$ the beam linear charge density,
and $E^{(0)}_n$ the (known)  field component 
normal to the pipe wall in the perfectly conducting pipe\footnote{
%+++++++++++++++++++++++++++++++++++++++++++++++
A similar formula exists for the (dyadic) transverse impedance  \cite{Partacc}.}.
%+++++++++++++++++++++++++++++++++++++++++++++++
%
According to (\ref{eq:ZZ}), perforated/slotted  or  OCMF patches
placed where  the normal field component $E^{(0)}_n$ is {\it minimum}
will have minimum impact on the longitudinal impedance.\\
Hence, a circular beam liner, featuring a {\it uniform} field along $\partial S$, 
represents a worst case, where the longitudinal (and transverse)
beam coupling impedances is simply proportional to the average of the
wall impedance, assumed as piecewise constant
\beq
\langle Z_{wall} \rangle
=\sum_i \xi_i Z_{wall}^{(i)}
\label{eq:aveZwall}
\eeq
$\xi_i$ being the surface fraction covered by patches with wall impedance $Z_{wall}^{(i)}$.  
%
%-----------------------------------------------------------------------------
\subsubsection*{Impedance of Perforated Patch}
%-----------------------------------------------------------------------------
%
\noindent
The effective wall impedance of a perforated patch can be computed 
under the Bethe approximation where the holes/slots 
are (much) smaller than the shortest wavelength in the beam spectrum,    
yielding  \cite{Kurennoy}, \cite{holes}
\beq
Z_{s}^{(0)}\!=\! \imath Z_0
\left(\frac{\omega}{c}\right)
\alpha_{tot}^{(i)} n_\sigma
\label{eq:uno}
\eeq
$Z_0\!=\!(\mu_0/\epsilon_0)^{1/2}$ being the vacuum characteristic impedance,
$\alpha_{tot}^{(i)}$ the total internal polarizability of the holes/slots,  
and $n_{\sigma}$ their surface density.\\
In \cite{Gluck} the more general case, relevant for the LHC,
of a circular perforated liner with inner radius $a$ and  thickness $\Delta$
surrounded by a circular co-axial lossy tube (the cold bore) 
with  radius $b$ was investigated.  
It was found that the liner wall impedance 
can still  be cast in the form of eq. (\ref{eq:uno}), 
after the formal substitution
\beq
\alpha^{(i)}
\longrightarrow
\alpha^{(i)}+
F
\alpha^{(e)},
\eeq
where the superfix $(e)$ identifies the total
{\it external} polarizability 
of the holes/slots  in  wall with thickness $\Delta$, and  \cite{holes}, \cite{Gluck}
\beq
F=-\frac{\alpha^{(e)}}{\alpha^{(i)}}
\left[
1\!+\!
\imath \left(1\!+\!\frac{b}{a\!+\!\Delta}\right)
\frac{Z_{cb}}{Z_{h}^{(0)}}
\right]^{-1},
\eeq
$Z_{cb}$ being the cold bore wall impedance \footnote{
%++++++++++++++++++++++++++++++++++++++++
In this formula $Z_{cb}$ is the impedance of both 
the the cold bore and outer beam-liner walls.}.
%++++++++++++++++++++++++++++++++++++++++
%
The wall reactance is negligibly affected by the presence of the 
external tube, which produces a small resistive component, 
accounting for radiation leakage throught the slots.\\  
We use the following formulas from for thin axial slots from \cite{KurKEK} :
\beq
\alpha_{tot}^{(0)} = w^3(0.1334-0.005 w/h), \mbox{  }
\alpha_{tot}^{(i)} = (8/\pi^2) \alpha_{tot}^{(0)}, \mbox{  }
\alpha_{tot}^{(e)} = \exp(-\pi\Delta/w) \alpha_{tot}^{(0)},
\label{eq:alphas}
\eeq 
where $\alpha_{tot}^{(0)}$ is the total polarizability of
a slot in a vanishingly thin wall, and
all other symbols are defined in Figure \ref{fig_5}.\\
%
%------------------------------------------------------------------
\subsubsection*{Impedance of OCMF Patch}
\label{sect:OCMF_imp}
%------------------------------------------------------------------
%
\noindent
A straightforward  solution,  which is not included for brevity, 
of the electromagnetic boundary value problem 
for a (relativistic, vanishingly thin) axial beam 
in a circular conducting foam liner, with radius $a$ and thickness $\Delta$
surrounded by a  co-axial (infinitely thick) conducting circular tube (the cold bore) 
with radius $b > a+\Delta$,  
shows that  if  $\Delta$ exceeds a few skin depths $\delta_f$
across the whole beam current spectrum,
which is certainly the case here, as seen from eq. \ref{eq:foam_delta},
the  Leont\'{o}vich impedance of the OCMF liner wall 
is fairly well approximated by the intrinsic impedance of the OCMF, viz., 
(compare w. eq (\ref{eq:foamZk}) 
\beq
Z_{f} =  Z_0
\left(
\frac{\nu}{2\omega_p}+
\imath \frac{\omega}{\omega_p}
\right).
\label{eq:foamZk2}
\eeq
The surface {\it roughness} of the foam also contributes 
to the OCMF wall impedance. The order of magnitude of this contribution 
can be estimated from \cite{BaStu_rough}  
\beq
Z^{(rough)}_{f}\approx
\imath
\sqrt{\frac{\pi}{32}}
Z_0 \left(\frac{h}{L}\right)
\frac{\omega}{c} h,
\label{eq:Z_rough}
\eeq
$h$ and $L$ being the  r.m.s. height  and correlation
length of the surface roughness, respectively.\\
%
%%%%%%%%%%%%%%%%%%%%%%%%%%%%%%%
\subsubsection*{Impedance Budget}
%%%%%%%%%%%%%%%%%%%%%%%%%%%%%%%
%
\noindent
For illustrative purposes, 
the numerical values of the real and imaginary components 
of the relevant wall impedances, normalized to the mode number
(i.e., multiplied by the $(\omega_R/\omega)$ factor),
have been collected in Table III.
Here we assume high-grade Copper 
($\rho\approx 5.5 10^{-10} \mbox{ ohm cm}^{-1}$ at $20K$) 
for the solid, slotted, and foam-matrix metal. \\ 
The numbers in Table III for the slotted-patch 
are obtained for the slot geometry depicted in Figure \ref{fig_5}, with
$w\!=\!\Delta\!=\!1.5mm$, $h\!=\!10w$ and $s_T\!=\!s_L\!=\!2w$, 
for which $\xi_h \approx 0.14$, and the escape probability
reaches its the fiducial upper limit $f \approx 0.1$. \\
The numbers for the foamed patch are obtained assuming 
a typical pore diameter $\sim 1\mbox{ mm}$, and a typical
ligament size $\sim 0.1 \mbox{ mm}$, yielding 
$\omega_p \approx 7.93 \cdot 10^{10} \mbox{ rad sec}^{-1}$  
and $\nu \approx 3.66 \cdot 10^4 \mbox{ Hz}$ 
in the Drude  model of Section \ref{sec:Drude}, 
consistent with typical measured values of 
the static conductance of open-cell Copper foams.  
We assume a typical r.m.s. roughness scale $h\approx .125 \mbox{ mm}$
and a correlation length $L \approx 0.25 \mbox{ mm}$.
%
%++++++++++++++++++++++++++++++++++++++++++++
\begin{table}[h]
\centering
\begin{tabular}{| c | c | c | c | c |}
\hline\hline
& 
$\mbox{ Solid Copper }$                             & 
$\mbox{ OCMF }$                           & 
$\mbox{ OCMF }$                          & 
$\mbox{ Slotted Patch }$\\
                                                     & 
$\mbox{Patch }$        & 
$\mbox{ Patch }$                           & 
$\mbox{ Roughness }$                          & 
$\mbox{ (Perfect Conductor) }$\\
\hline
$(\omega_R/\omega) R_{wall}, \mbox{ [ohm]}$    &   
$4.9\cdot 10^{-6}  \sqrt{\omega_R/\omega}$  &     
$1.4 \cdot 10^{-5}  (\omega_R/\omega)$            &
$0$     &       
$1.0\cdot 10^{-13} \sqrt{\omega/\omega_R}$ 
\\
\hline
$(\omega_R/\omega) X_{wall}, \mbox{ [ohm]}$)     &      
$4.9\cdot 10^{-6}  \sqrt{\omega_R/\omega}$   &     
$5.5 \cdot 10^{-5}$ & 
$5.5 \cdot 10^{-6}$ & 
$3.1 \cdot 10^{-7}$   
\\
\hline\hline
\end{tabular}
\caption{ \it Summary of Leont\'{o}vich impedances. 
High-grade Copper with $\rho\approx 5.5 10^{-10} \mbox{ohm cm}^{-1}$ at $20K$. 
OCMF with $1 \mbox{mm}$ pores and $0.1  \mbox{mm}$ ligaments.
Slots in a perfectly conducting wall  as in Figure \ref{fig_5},  with 
$w\!=\!\Delta\!=\!1.5mm$, $h\!=\!10w$ and $s_T\!=\!s_L\!=\!2w$.}
\label{table3}
\end{table}
%++++++++++++++++++++++++++++++++++++++++++++
%
\\
As seen from Table III,  
the Copper wall resistance is larger than that of the OCMF
up to  $\omega=10^5\omega_R$.
On the other hand, as anticipated in Section \ref{sec:Drude}, 
the  OCMF wall reactance is relatively large, 
and exceeds significantly that of a perfectly conducting slotted wall
as well as that of solid Copper.\\
However, as seen from Figure \ref{fig_5}, a typical OCMF wall pumping capacity 
(escape probability) is several times larger than that of the
''best'' slotted wall.
Thus, to obtain the same pumping capacity, the patched
beam-liner surface  needed in the OCMF case is only a small fraction
of that for the slotted case.  
This makes easier to place  the OCMF patches where the (unperturbed) field 
in eq. (\ref{eq:ZZ}) is minimum, so as to minimize their impact 
on the beam coupling impedance.\\
%
%For example, according to Figure \ref{fig_5},
%for a beam liner with $f=0.1$,  
%which is the fiducial upper limit for a slotted patch with
%$w\!=\!\Delta\!=\!1.5mm$, $h\!=\!10w$ and $s_T\!=\!s_L\!=\!2w$,
%one needs to cover the {\it entire} pipe surface using slotted patches,
%and only a fraction $\approx .125$ of the pipe surface using OCMF patches 
%with $p=0.9$ and $\Delta=2 \mbox{mm}$.
%
%+++++++++++++++++++++++++++++++++++++++++++++++++++
\subsection{Secondary Emission Yield}
\label{sec:SEY}
%+++++++++++++++++++++++++++++++++++++++++++++++++++
%
Perforations reduce the secondary emission yield  (SEY)  
of a metal wall by a factor roughly proportional
to the solid fraction of the surface  \cite{roughSEY}.\\
The SEY  of metals is strongly reduced by surface roughening,
obtained, e.g., by powder blasting \cite{ecloud2}.
For the special case of periodic rectangular grooves  etched on a flat metal surface,
the SEY dependence on the groove depth-to-spacing and  width-to-thickness ratios  
has been investigated, indicating that larger ratios yield a smaller SEY \cite{Pivi}.\\
Carbon coating also reduces  SEY effectively  \cite{Carbon},  at the expense
of some increase in wall resistance.\\
It is reasonable to expect that OCMFs  
may exhibit a much lower SEY compared to solid metals
in view of their random-like surface roughness.
The SEY of an OCMF wall will depend on the effective roughness
of its surface, which is a function of the foam porosity 
and  average  pore size.  
OCMF surface roughness, however, should be kept small enough, 
to prevent the blow up of wall rectance.
No measurements of the SEY of OCMFs are available yet,
to the best of our knowledge, but should be straightforward \cite{SEYmeasu} .
%
%+++++++++++++++++++++++++++++++++++++++++++++++++++
\section{CONCLUSIONS}
\label{sec:conclu}
%+++++++++++++++++++++++++++++++++++++++++++++++++++
%
On the basis of the above discussion, some tentative conclusions can be drawn 
about the possible use of OCMF materials in the
beam liners of high synchrotron radiation accelerators.\\
The outgassing capabilities of OCMF patches can be better 
by a factor $\sim 10$ compared to the ''best'' slotted patches, 
and would accordingly allow to boost the molecular escape probability
of beam pipes by roughly one order of magnitude, 
compared to present designs,
while keeping the patched fraction of the beam-liner surface unchanged.\\
The mechanical-structural properties of conducting foams should also be adequate
to resist to eddy-current induced stresses, 
in view of their very morphology.\\
A low SEY is also expected, due to OCMF surface roughness, even though
no experimental results are available yet.\\
The OCMF surface resistance is very low, and almost frequency independent.
On the other hand, the (inductive) wall reactance of an OCMF patch 
is larger by roughly one order of magnitude compared to that of a perforated 
wall with comparable escape probability. 
This drawback could be mitigated by clever placement of the OCMF
patches, and/or appropriate reactive loading 
of the (solid portion of) the pipe wall. \\
In order to translate the above hints into effective design criteria,  
further modeling effort and substantial experimental work are in order.
Measurement of  the complex surface impedance and SEY of  metal foams 
are now underway \cite{Cimmin}.\\
We believe that such an effort is worth being pursued, 
and that the available modeling tools and technologies provide a good starting point 
for its successful implementation.
%
%+++++++++++++++++++++++++++++++++++++++++++++++++++
\section*{Acknowledgements}
%+++++++++++++++++++++++++++++++++++++++++++++++++++
%
This work has been sponsored in part by the Italian Institute for Nuclear Physics (INFN) under the IMCA grant. Stimulating discussions  with F. Zimmermann and G. Rumolo (CERN) are gratefully acknowledged. 

\end{document}